\documentclass[english]{article}
\usepackage[utf8]{inputenc}
\usepackage[T1]{fontenc}
\usepackage{babel}
\usepackage{amsmath}
\usepackage{graphicx}
\usepackage{fancyhdr}
\usepackage{hyperref}
\usepackage[printwatermark]{xwatermark}
\usepackage{xcolor}

\begin{document}

\title{The Zoltar forecast archive: a tool to facilitate standardization and storage of interdisciplinary prediction research}

\author{
Nicholas G Reich\textsuperscript{1{*}}, 
Matthew Cornell\textsuperscript{1},
Evan L Ray\textsuperscript{1}, \\
Katie House\textsuperscript{2},
Khoa Le\textsuperscript{2}
}

\maketitle
\thispagestyle{fancy}

\noindent 1. Department of Biostatistics and Epidemiology, University of Massachusetts-Amherst, Amherst, MA, 01003, USA\\
\noindent 2. College of Information and Computer Sciences, University of Massachusetts-Amherst, Amherst, MA, 01003, USA\\
{*}corresponding author: nick@schoolph.umass.edu

\begin{abstract}
Forecasting has emerged as an important component of informed, data-driven decision-making in a wide array of fields.
We introduce a new data model for probabilistic predictions that encompasses a wide range of forecasting settings.
This framework clearly defines the constituent parts of a probabilistic forecast and proposes one approach for representing these data elements. 
The data model is implemented in Zoltar, a new software application that stores forecasts using the data model and provides standardized API access to the data.
In one real-time case study, an instance of the Zoltar web application was used to store, provide access to, and evaluate real-time forecast data on the order of 10$^7$ rows, provided by over 20 international research teams from academia and industry making forecasts of the COVID-19 outbreak in the US.
Tools and data infrastructure for probabilistic forecasts, such as those introduced here, will play an increasingly important role in ensuring that future forecasting research adheres to a strict set of rigorous and reproducible standards.


\end{abstract}


\section*{Introduction}


Standardized data curation has emerged as a central challenge in many scientific disciplines.
As public awareness of the so-called ``reproducibility crisis'' in scientific research has grown, so has the recognition of the importance of standardized structures and repositories for research data.
Improving practices of dataset registration, data sharing, and standardizing definitions have been identified as key factors that could lead to improving the robustness of scientific results.\cite{ioannidis2014make}

Over the past few decades, forecasting has emerged as a resource for supporting decision-making in a wide array of fields. 
Much early forecasting literature was devoted to applications in economics \cite{Roos1955,Granger1973,Diebold1998} and meterology \cite{Murphy1984,Kalnay1996}.
However in recent years, other fields such as energy \cite{Hong2014,Hong2016,Debnath2018}, epidemiology \cite{biggerstaff2016results, viboud2017rapidd}, demography \cite{gerland2014world,Alkema2015,Azosea2016}, politics \cite{campbell1996polls, graefe2015german, lewis2014us}, sports \cite{Stekler2010,Santos-Fernandez2019}, and seismology \cite{field2009uniform,Chambers2012} have also turned to forecasts to support real-time decision making and to improve situational awareness.

Some research groups and consortia have developed discipline-specific databases of forecast data, to create findable and accessible versions of these data.
\begin{itemize}
    \item The European Centre for Medium-Range Weather Forecasts (ECMWF) is a research institute that provides 24/7 operational weather forecasts (\href{https://www.ecmwf.int/}{https://www.ecmwf.int/}). The ECMWF has multiple publicly available datasets available for reanalysis and reuse (\href{https://apps.ecmwf.int/datasets/}{https://apps.ecmwf.int/datasets/}). 
    \item The Subseasonal Experiment (SubX) is an experimental project at the US National Oceanic and Atmospheric Administration (NOAA) (\href{http://cola.gmu.edu/kpegion/subx/}{http://cola.gmu.edu/kpegion/subx/}) that provides experimental, real-time weather forecasts each week. SubX data are available from an online repository in a standardized format.\cite{subx} 
    \item  The FiveThirtyEight data journalism website stores forecasts of sports and political events with fixed 0/1 outcomes. The data on over 1.5m forecasts of binary event outcomes (i.e. elections and sports games) are available in a single plain text file on GitHub under a CreativeCommons license.\cite{538forecastdata} Additionally, they have conducted detailed analyses examining the performance of their forecasts.\cite{538forecasteval}
    \item The US Centers for Disease Control and Prevention (CDC) store over 1.8GB of data for real-time forecasts of influenza seasons since 2015 on GitHub. These forecasts are available under a Creative Commons license.\cite{cdcforecastdata}
\end{itemize}
These existing datasets are noteworthy for their public accessibility. 
However, they lack a shared common set of cross-disciplinary definitions and standards. 

It is against this backdrop that we introduce Zoltar, a forecast repository, as a data curation tool to improve the robustness of forecasting research.
Zoltar is a research data repository server that stores forecasts made by external models in standard formats and provides tools for validation, visualization, scoring, and forecast registration or time-stamping. 
It builds on a foundation of core ideas and data structures introduced in 2019 by {\tt predx}, a software package that defines formal data structures for different kinds of predictions.\cite{predx}


In recent large-scale collaborative forecasting research projects, the need for a system to consolidate, standardize, and facilitate the reuse of forecast data has become abundantly clear.\cite{ george2019technology} 
In one example, a GitHub repository maintained by a subset of authors of this article houses nearly 5GB and over 30m rows of highly structured probabilistic forecast data used in multiple scientific publications about influenza forecasting.\cite{reich2019accuracy,reich2019collaborative} 
However, these data have been painstakingly manually curated, without integrated data validation, programmatic access for external collaborators, or standardized code for evaluating forecast accuracy.
While there is no specific reason to believe that errors have been committed in maintaining this repository, a more structured approach to data curation would provide additional assurance and allow for such a structure to be more easily scaled and leveraged by a wider variety of research groups.

Our work on Zoltar has generated two important new developments for forecasting research.
First, we define a new data model for representing probabilistic forecasts. 
Probabilistic forecasts are an increasingly common feature of forecasting projects and research.\cite{Winkler2019} 
Some efforts have been undertaken to create object-oriented representations of probability distributions in specific programming languages.\cite{sonabend2020}
However, to our knowledge, there is no agreed-upon set of general standards for representing different kinds of probabilistic forecasts.
The current work presents a coherent data model and one possible storage structure for probabilistic forecasts. 
Second, we have developed software for a hosted application on a web server. 
Instances of this web application, such as the operational one, Zoltar, at \href{https://zoltardata.com}{zoltardata.com}, use this new probabilistic forecast data framework to curate real-time forecasting datasets.
This system allows for forecasts to be registered with a timestamp at the time they were uploaded, easily accessed via an API, and evaluated using standard metrics.

The central goals of Zoltar are to 
\begin{itemize}
    \item provide an open-source data curation framework for forecasting projects
    \item formalize and standardize a general data model for forecasts
    \item make structured forecast data accessible programmatically
    \item facilitate comparisons of forecast accuracy within and across disciplines, and over time
    \item provide a time-stamp registration for forecasts of future events
    \item provide standardized, interactive tools for scoring and visualizing forecasts
\end{itemize}

The Zoltar forecast archive is available at \href{https://zoltardata.com/}{https://zoltardata.com/} with detailed documentation available at \href{https://docs.zoltardata.com/}{https://docs.zoltardata.com/}.

\section*{Results}

\subsection*{A new data model for probabilistic forecasts}
\subsubsection*{Probabilistic forecasting overview}
Relative to other scientific datasets, forecast data have some unique features.
A forecast can be defined as a quantitative statement about an outcome variable that has not yet been observed, conditional on data that have been observed.
Forecasts are often probabilistic, meaning they assign probabilities to all possible outcomes by specifying a valid analytical or empirical probability distribution.
Probabilistic forecasts are increasingly common \cite{Winkler2019} and also have additional complexity beyond that of point forecasts in terms of data structure and storage space required.

\subsubsection*{Forecast metadata}
\begin{figure}[h]
    \centering
    \includegraphics[width=\linewidth]{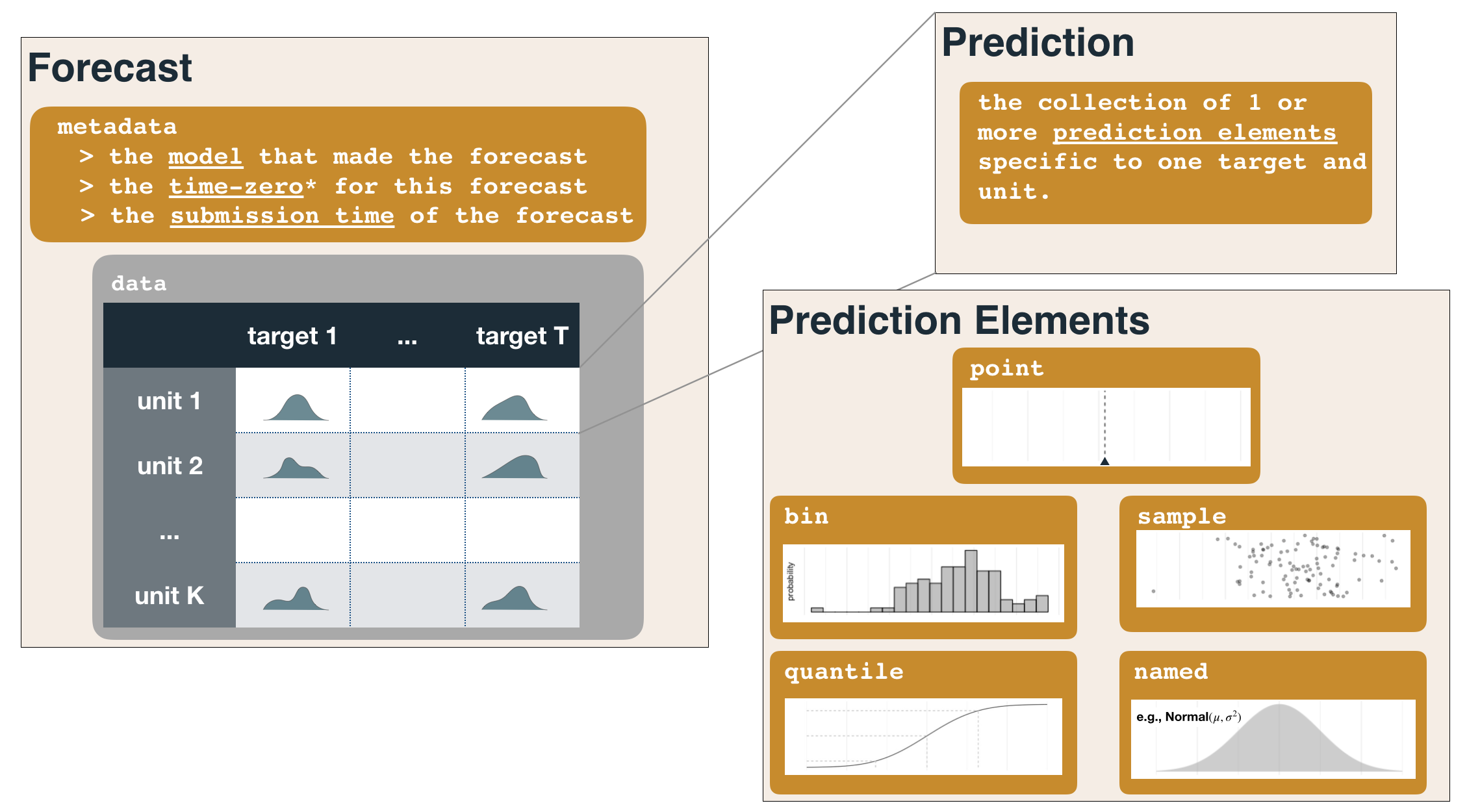}
    \caption{Overview of the structure of the data contained within a forecast. A forecast contains metadata and forecast data (represented by the shaded curves) for each of a set of units and targets. A prediction, in our nomenclature, refers to the set of forecast data for one submitted forecast, and one target-unit pair. Each prediction is made up of one or more prediction elements.}
    \label{fig:forecast-schematic}
\end{figure}

At its core, a forecast is a quantitative representation of prediction(s) for one or more targets of interest. 
However, it is also important to record additional structured metadata to contextualize a forecast.
Our proposed data model for forecasts includes quantitative representations of predictions for combinations of particular units (e.g., different geographical units) and forecasting targets, as well as a set of metadata about the model (Figure \ref{fig:forecast-schematic}).

Forecast metadata includes information about the model that made the forecast and the time at which the forecast was registered in the system.
This \mbox{``timestamp''}, if appropriately and independently registered, can help ensure that the forecast was indeed made prior to the event of interest being observed. 

Additionally, we use the concept of a `time-zero', which is the the temporal reference point for which any step-ahead forecasts are in reference to.
`Time-zeros' also serve as pre-specified slots, identified by a date (although more generally this could be an exact time), for which forecasts are made. 
Any model can only have one forecast associated with each `time-zero'.
Therefore, a `time-zero' is included as metadata for each forecast (Figure \ref{fig:forecast-schematic}). 
For example, a ``1-step ahead'' forecast for a particular time-series is one time step (e.g., a day, week, or month) in the future relative to the time-zero date.

\subsubsection*{Representations of forecasts}

To allow for consistent and standardized representations of probabilistic forecasts, we developed a data model for forecasts.
In the data model, the quantitative information about each specific prediction can be represented by five different data structures.
\begin{itemize}
    \item A {\bf Named} representation is a probabilistic forecast that follows a known probability distribution. A distribution family is specified, along with the parameters for that distribution.
    \item A {\bf Sample} representation is a probabilistic forecast represented by empirical samples from a distribution.
    \item A {\bf Bin} representation is a probabilistic forecast represented by an empirical distribution. The set of possible values of the quantity being predicted is divided into pre-specified bins, and each bin is assigned a predicted probability.
    \item A {\bf Quantile} representation is a probabilistic forecast represented by a series of quantiles of the distribution.
    \item A {\bf Point} representation is a forecast providing a single predicted value for the outcome of interest.
\end{itemize}
In practice, the above list moves from more detailed and explicit (at the top) to more coarse and less informative (at the bottom). 
A Named distribution could be unambiguously translated into either a Sample, Bin, or Quantile representation; 
a Sample representation could be translated into a Bin or Quantile representation;
and any of the three probabilistic representations could be used to provide a point estimate. 
Depending on the resolution of Bin or Quantile representations, approximate Sample representations could also be derived.
See Methods for further details about the specifications and storage of these different types of predictions.

\subsubsection*{Types of forecasting targets}

Additionally, our data model specifies different types of variables that are the ``targets'' for a particular forecasting effort.
\begin{itemize}
    \item {\bf continuous}: A quantitative target whose range is an interval subset of the real numbers. Examples: percentage of all doctors' office visits due to influenza-like illness, or disease incidence per 100,000 population.
    \item {\bf discrete}: A quantitative target whose range is a set of integer values.  Example: the number of incident cases of a disease in a given time period.
    \item {\bf nominal}: A categorical target.  Example: severity level in categories of "low", "moderate", and "high", or the winner of an election "candidate A", "candidate B", or "candidate C".
    \item   {\bf binary}: A binary target, with a defined outcome that can be seen as a true/false. Example: does a stock market index exceed some threshold C by a specific date, or the winner of a two-team sports match.
    \item {\bf date}: A target with a discrete set of calendar dates as possible outcomes. Example: the day on which peak disease incidence occurs in a given period of time.
\end{itemize}
Different prediction elements are available for different target types (Table \ref{tab:target-types}).
Targets are defined in greater technical and mathematical detail in the Methods section.

\subsection*{Zoltar feature set}

The following sections outline the structure and feature set of the Zoltar forecast repository. Complete documentation of the Zoltar system can be found at \href{https://docs.zoltardata.com/}{https://docs.zoltardata.com/}.

\subsubsection*{Structured forecast data storage and management}

\begin{figure}
    \centering
    \includegraphics[width=\linewidth]{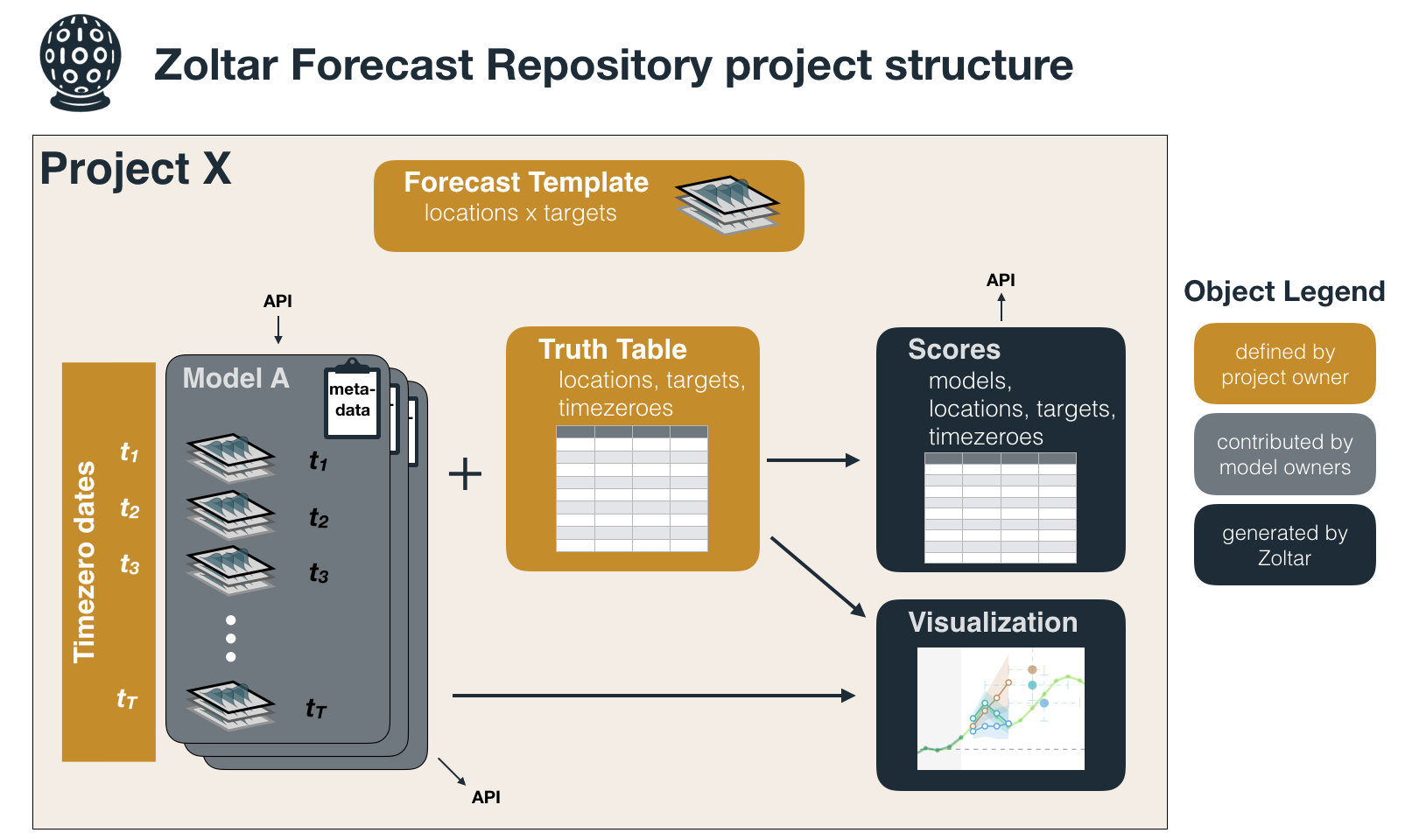}
    \caption{Overview of the Zoltar Forecast Repository structure. The repository is made up of projects. Project owners define a template for a forecast by specifying units and forecast targets. They also specify the times for which forecasts should be made. They can optionally specify a ``Truth Dataset'', containing eventually observed values for each target. Zoltar users designated as ``Model owners'' for the project can upload forecasts associated with their model via a web user interface or RESTful API (see Methods). These forecasts are run through data validation checks before entering the system. If a truth dataset is present, Zoltar generates data visualizations and scores the forecasts.}
    \label{fig:overview}
\end{figure}

Zoltar uses a project-based organizational structure (Figure \ref{fig:overview}). 
A project contains all the information needed to delineate the scope of a particular forecasting exercise. 
Any Zoltar user may create a project, and as the project owner they specify sets of units, targets, and time-zeroes.
Projects may be designated as public (data are readable by anyone accessing the site) or private (only specified registered users have read access to the data).

Models are then associated with projects and registered Zoltar users who are designated as ``model owners'' for a particular project can upload forecasts for their model when they are created.
Project owners can upload (and update, as needed) the eventually observed values associated with the units, targets, and time-zeroes for each project. 
When such a truth dataset is uploaded, Zoltar automatically generates a set of standard scores and metrics for all submitted forecasts (Table \ref{tab:scores}).
Forecasts can be uploaded to and downloaded from Zoltar using a RESTful API (see details next section).
Scores can also be downloaded using the API.

\subsubsection*{API access}

All data elements in Zoltar can be browsed and queried through a RESTful API framework.
Users may browse the website, and, with a click on the ``API'' button, have access to the core data for each of the core data elements: projects, models, and forecasts.
Retrieval of data through the API is facilitated by the {\tt zoltr} package for R and the {\tt zoltpy} library for Python, which retrieve data in the native data structures of each programming language.\cite{cornell2020zoltr,cornell2020zoltpy}
When necessary for data access, authentication is enabled via JSON Web Token (JWT).
All project details, such as units, targets, and time-zeros, can also be managed via calls to the API.

\subsubsection*{Data validation}

All data in Zoltar goes through data validation checks to ensure consistency and validity of the data.
Forecasts are validated upon upload, and warning or error messages are sent to the model owner if a forecast does not pass a validation test.

Validation tests are specific to the type of forecast in Zoltar. 
The online documentation provides details on the forecast validations that are in place. 
At the time of writing, up to 47 tests are run when each forecast is uploaded to Zoltar. 
For example, all predictions are verified to ensure that the data type of incoming data must be valid for its target's type (e.g. a floating point number cannot be submitted for a date target).

\subsubsection*{Forecast scoring}

As described above, for projects that contain both a truth dataset and forecasts, Zoltar generates scores automatically (Table \ref{tab:scores}).
As of writing, Zoltar computes the residual error and absolute error for point forecasts, the log score and probability integral transform for bin and sample forecasts, and the interval score for quantile forecasts.\cite{Gneiting2007} 
Brier scores and continuous ranked probability scores (CRPS) are planned additions.
Aggregate scores may be viewed on the website, and individual forecast scores are available for download on a project-by-project basis through the website user interface or programmatically via the API.

\subsubsection*{Forecast visualization}

\begin{figure}
    \centering
    \includegraphics[width=\linewidth]{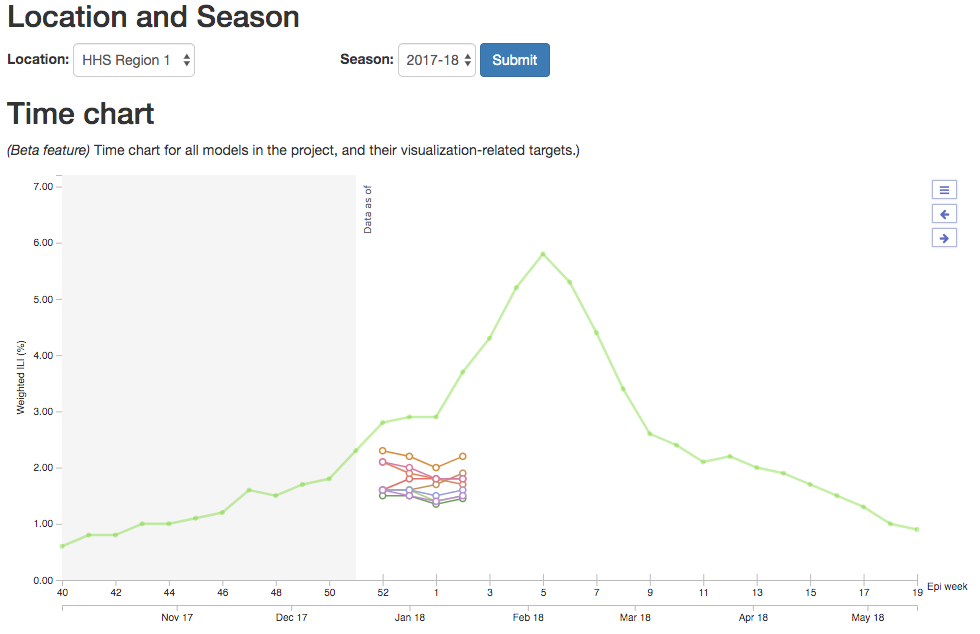}
    \caption{Screenshot of visualizations of Zoltar forecast data. Ground truth data is shown by the green data in the background. Point forecasts for various models are shown (colored dots and lines).}
    \label{fig:viz-example}
\end{figure}

The {\tt d3-foresight} javascript library is an interactive data visualization toolkit specifically designed for visualizing time-series forecasts.\cite{tushar2017flusight}
Zoltar uses {\tt d3-foresight} and the available forecast data to generate interactive visualizations for time-series forecasting projects (Figure \ref{fig:viz-example}).
Project owners must specify project-specific settings for the visualization, e.g., the scale for the forecasted variable, and which targets are ``step-ahead'' targets.
Step-ahead targets are ones that are future values of a given time-series. For example, if the forecasted time series is the number of weekly incident cases of influenza, then the step-ahead targets for time-zero of $t$ are the number of cases in week $t+1$, $t+2$, etc.

\subsection*{Forecast archive use cases}

\subsubsection*{Templates for new research and forecasting challenges}

Zoltar's open, documented data structures provide users interested in initiating new research projects or forecasting challenges with a set of tools and templates.

For example, if a non-governmental organization concerned with health-related issues wanted to organize a forecasting challenge similar to the CDC FluSight challenge\cite{biggerstaff2016results}, they could download the existing project template file, modify it for their purposes, and create a new project to solicit forecasts for their setting. 
Zoltar significantly lowers the technical barriers that currently exist for creating and managing such a challenge, by standardizing the required data input formats and also automatically calculating multiple scores for each target.
While the functionality offered by Zoltar here is somewhat similar to that offered by a prediction competition websites, such as Kaggle.com, its focus on data formatting standards for probabilistic outcomes, the specification of a formal data model for forecast data, and features of standardized scoring and visualization provide additional functionality above and beyond the basic Kaggle competition structure.

\subsubsection*{Teaching and workshops}

An instructor in forecasting could design a Zoltar project for a class or a workshop where the final deliverable for a project is a set of forecasts.
With the suite of tools and data structures available to facilitate this process, an instructor would not need to spend as much time developing standardized data formats or writing code to parse, store, and score a variety of forecasts from different groups. 
If desired, a ranking of the scores of submitted forecasts could be easily downloaded from Zoltar soon after forecasts were uploaded.

\subsubsection*{Retrieving data for analysis and visualization}

For Zoltar projects with open-access data, datasets could be used as inputs into new, externally developed visualization tools.
Efforts to develop such tools have been impeded by lack of standardized data repositories and access.\cite{george2019technology}
Whether for professional, research, or hack-a-thon projects, Zoltar provides a set of rich and scientifically interesting data sources in a standard and accessible format. 

\subsubsection*{Comparing forecasts to each other and benchmark models}

Conducting research on how new or existing methods perform in specific forecasting application settings will be facilitated by Zoltar. 
A researcher could create a new project on Zoltar designed around a particular set of targets of interest. 
Either on her own or in collaboration with other research groups, the researcher could populate the project with forecasts from a wide array of existing and new models. 
In many ways, the analysis of Reich et al. (2019), which served a motivator for developing Zoltar, is a prototype of this kind of analysis.\cite{reich2019collaborative}
In this work, a reference baseline model was identified, and an array of forecasts from other state-of-the-art modeling approaches were compared in a standardized way.
This was the first time in this particular research community that performance comparisons across multiple years were able to be made between models from different groups. 

Such structures encourage honest and robust evaluation of different methodologies. 
By making it simpler to compare a new forecasting method to existing methods, standardized comparisons with existing methods will be easier to generate. 
For example, even after a paper like the one described above has been written, a new group with a new methodology could come along and upload a set of forecasts for comparison with the existing data.
In this way, Zoltar can serve to build community standards and benchmarks of forecasting studies across a range of disciplines.

\subsubsection*{Building ensemble models}

In many forecasting applications, the ultimate goal is to create an ensemble forecast that combines forecasts from multiple different models.
Multiple features of Zoltar have been designed explicitly to assist with this task.
First, verified forecasts are available in standard, documented formats. This facilitates combining multiple forecasts, without having to reformat or check for formatting errors.
Second, forecasts are available for automated download via a queryable API.
Third, multiple verified scores for available forecasts are available for additional training of ensemble parameters, if needed.\cite{reich2019accuracy}
In May 2020, the COVID-19 Forecasts project on Zoltar was used for this purpose, with forecasts downloaded from Zoltar used to create the COVID-19 Forecast Hub ensemble model displayed on the US CDC website.\cite{CDC2020} 

\subsection*{Case study: forecasts for COVID-19}
Beginning in mid-March 2020, our team led a collaborative effort to bring together forecasts of deaths due to COVID-19 in the US.\cite{covidforecasthub} 
The forecasts from this ``COVID-19 Forecast Hub'' project are being automatically uploaded to and stored in a project on the Zoltar website.\cite{Zoltar-covid-project} 
In the first 6 weeks of the project (through May 31, 2020), over 17m rows of forecast data were added to the project from 45 models.
This project contains quantile and point forecasts of the following targets with public-health relevance:
\begin{itemize}
    \item 1 through 20 week-ahead cumulative deaths
    \item 1 through 20 week-ahead incident deaths 
    \item 0 through 130 day-ahead cumulative deaths
    \item 0 through 130 day-ahead incident deaths
    \item 0 through 130 day-ahead incident hospitalizations
\end{itemize}
Forecasts were accepted for the US national level and the 57 states, territories, and jurisdictions. 
Data stored in this Zoltar project were the official source of the data on the CDC COVID-19 Forecasts website beginning in early April 2020.\cite{CDC2020}


Each week, recent forecasts were downloaded through the Zoltar API to create a weighted average ensemble forecast, a combination of the forecast data from all relevant models.
This downloading process took advantage of the Zoltar feature that allows for querying the API to only return certain subsets of forecasts.
Forecasts from this ensemble model (also stored in Zoltar as the {\tt COVIDHub-ensemble} model) was presented on the US CDC website, as were forecasts from individual models also stored in the repository.\cite{CDC2020} 

\section*{Dicussion}

The amount of published research on forecasting has doubled in each of the past two decades.\cite{lauer2020}
As forecasting assumes a more prominent role in scientific inquiry in many disciplines, new guidelines are needed to ensure that forecasting research follows rigorous standards.\cite{pollett2020}
Additionally, new technology and data architecture are needed to ensure interoperability and comparability between different forecasting systems.\cite{george2019technology}
The forecast data model and online archive introduced by this manuscript were designed to address both of these needs. 

In this work, we introduce Zoltar, a system that we hope will serve the needs of forecasters in academia, government, and industry.
Our data model provides a starting point for thinking carefully about representations of probabilistic forecasts that will be useful in different contexts. 
Zoltar lowers the barriers to creating forecasting exercises that can help accelerate knowledge discovery in different fields.\cite{Lutz2019}

Forecasting is a unique field of research because of its direct link to operational decision-making in fields as varied as stock market trading, weather, energy, infectious disease epidemics, sporting events, and politics.
As a result, the range of forecasting efforts in different fields and for different audiences is astoundingly broad.
Operational forecasts with professional data pipelines and regular updates, such as weather forecasting platforms or some mass media political forecasting, by virtue of their mass market dissemination, command large audiences.
However, the degree to which these models are honestly validated in retrospect varies quite substantially.
There are nice examples of forecast evaluation in weather\cite{floehr2010weather} and politics/sports\cite{538forecasteval} forecasting.
By providing an open and transparent data structure, we hope to increase the number of rigorous academic forecasting studies that are conducted in real time.
Zoltar provides standardized data templates and storage structures, and will contribute to lowering the barriers to conducting large-scale forecasting projects and research.

There are many unique challenges in conducting forecasting research, and some can be addressed by developing a common set of data standards.
Better data infrastructure, like the services offered by Zoltar, can support the development and establishment of domain-specific benchmark models. 
For real-time forecasting studies, independently verifiable time stamps ensure that no forecasts had the benefit of being made after an event occurred.

However, other challenges in forecasting research exist indepenent of the data infrastructure available to researchers.
For retrospective studies, a particularly challenging problem is ensuring the new models generate forecasts as if they were operating in real time. 
This is challenging, because in some disciplines, forecasting models commonly rely on data that may be revised after it is first reported. 
For example, data on economic indicators or from epidemiological surveillance surveillance systems, are commonly revised based on updated information. 
(By contrast, such revisions are not typically present in meterological datasets.)
Repositories for economic\cite{croushore2001real} and epidemiological\cite{DELPHI} data have been set up for curating data with appropriate `vintages' and can serve as important complementary data infrastructure to tools like Zoltar.
Additionally, formal data structures and pipelines for the forecast modeling process itself could help pave the way for simple models to be created as benchmarks in a wide range of different application areas. 
Such pipelining efforts are in place for other types of analyses, but have not been designed or extended for the context of forecasting yet.\cite{Lang2019}

The data model and open-source tool presented in this manuscript will facilitate future development and standardization in multiple fields where forecasting is common.
While we believe that the tools presented here provide a solid foundation for growth, there are clear places where continued development is needed.
In this version of Zoltar, we have been primarily concerned with one-dimensional targets, i.e., targets where the eventual observation is a single value. 
Targets can be made functionally multivariate through the use of Sample-based representations of forecasts, which can encode covariance between multiple units and/or targets.
However, a more thorough understanding and detailed representation of multivariate targets should be explored.
Additionally, a natural extension to the work here would be to build tools to facilitate multi-model ensemble building and forecasting.


Forecasting research and practice is accelerating as the availability of standard data and usability of computational tools increase.
The development of a clear set of data standards for probabilistic forecasts has the potential to enhance and standardize the interoperability and evaluation of forecasting systems and frameworks across disciplines.
Tools and data infrastructure for probabilistic forecasts, such as those proposed here, will play an increasingly important role in ensuring that future forecasting research adheres to a strict set of rigorous and reproducible standards.

\section*{Methods}


\subsection*{Components of a forecast}

In common forecasting applications, an individual forecast will have a defined data structure (Figure \ref{fig:forecast-schematic}). These features include metadata about the forecast, targets, units (or locations) for which forecasts are being made, and data representing the forecasts themselves. We provide details on each of these features in the following subsections.

\subsubsection*{Metadata}

Every forecast has a set of meta-data associated with it, including the following items.

\begin{itemize}
    \item {\bf Model} (required): Every forecast must identify what model it was generated by.


    \item {\bf Time-zero} (required): The date from which a forecast originates and to which targets are relative (i.e. if a target is defined as ``the number of new cases two weeks after the current time'' then the time-zero is this idea of the ``current time''). Every forecast must have a time-zero associated with it. 

    \item {\bf Data version date} (optional): If the data version date exists, it is associated uniquely with a time-zero. It refers to the latest date at which any data source used for the forecasts should be considered. If present, it can be used externally by users retrospectively to recreate model results by "rolling back" the versioned data to a particular state.

    \item {\bf Time-stamp} (required): The time at which a forecast was formally registered in a public system. 
\end{itemize}

{\em Example:} As one example of when a forecast time-zero may be different than the time-stamp, consider the case of submitted forecasts for the CDC FluSight challenges. 
Forecasts that use official public health surveillance data collected up through week $t$ are typically submitted on Monday evening of week $t+2$. 
(Weeks are defined using the standard of MMWR weeks, where the first day of a week is Sunday.\cite{mmwr,Niemi2015,Tushar2018})
In this setting, the time-zero is defined as the Monday of week $t$, although forecasting models may use additional data available from other sources available to them up through the time of submission in week $t+2$. 
Then, 1- through 4-week ahead forecasts submitted for a time-zero in week $t$ of a particular target refer to weeks $t+1$ through $t+4$, respectively.

\subsubsection*{Targets}

Targets can be specified as different types of measured variables (Table \ref{tab:target-types}). 
Each target type has a data type (float, int, text, boolean, or date) associated with it, and can have certain kinds of forecasts made for it.

Let $y_t$ be an observed value of a (possibly multi-dimensional) variable in time interval $t$ from a time series $\{y_1, y_2, y_3, \dots\, y_t, \dots, y_T\}$.
We let $T$ refer to the total number of time points in a time-series and $t$ refer to a specific time-zero.

A forecast contains quantitative statements about forecast `targets', which are values or summaries of our measured variable of interest, $y$. 
Within a given forecasting exercise or project, there will exist a fixed number of specified targets. 
We will refer to the $i^{th}$ of these targets as $z_{i|t}$, target $i$ positioned relative to time $t$.
A target $z_{i|t}$ could be the value of a measured variable of interest (e.g., the value of a stock market index fund, the number of cases of a particular infectious disease, or the location in space of the eye of a hurricane) at subsequent time-steps, i.e., $z_{i|t}=y_{t+h}$.
A target could also be a summary statistic about a measured variable. For example, the first time at which the measured variable crosses a threshold value denoted by the constant $C$
($z_{i|t}= \min_{ \left \{ t': y_{t'} > C \right \} } ( t' )  $), 
or the highest value of the measured variable within a particular set $\mathcal{S}$ of times
($z_{i|t}= \max_{\left \{ t'\in \mathcal{S} \right \} } (Y_{t'}) \} $).

\subsubsection*{Forecast units}
Many forecasting applications will require multiple targets to be forecasted for multiple different units or locations.
Therefore, a forecast must have unambiguous, codes defining each location or unit.
Units could be different geo-political locations (states, countries, counties) or other organizational units (companies, electoral candidates, locations within a genomic sequence, etc...).

\subsubsection*{Predictions and prediction elements}

Forecasts of a particular target may be generated by models quite differently. 
One model might obtain a closed-form analytic expression for a probabilistic forecast.
Another might use samples from a posterior distribution to define an empirical distribution.
We have defined specific data structures, ``prediction elements'', that can be used to represent forecasts (see Figure \ref{fig:forecast-schematic} and details in the next section).
The five types of prediction elements are Named, Sample, Bin, Quantile, and Point.

\subsection*{Zoltar database structure}

\subsubsection*{Summary}

Conceptually, the structure of Zoltar was described above, and is shown in Figure \ref{fig:overview}.
Zoltar implements our data model using a standard relational database model, with the inner structure of Zoltar data represented in several core tables.
Each type of forecast has a corresponding table in the database. 
There is a single table for all point forecasts, another table for all bin forecasts, and so on. 
A separate set of metadata tables organize information about projects, and the associated targets and units.


\subsubsection*{Forecast data model}

In Zoltar, a ``forecast'' corresponds to a particular set of data associated with a unique combination unit and time-zero. 
Within a forecast, we say that a ``prediction'' corresponds to the data associated with one specific target.
And within the data for that particular target, there may be multiple different ``prediction elements''. 
For example, a prediction for the number of cases of a particular disease at a date in the future could be represented by a point forecast and one or two of the probabilistic forecast types (quantile, bin, named, or sample). 
Future development will allow for simple internal translations between the forecast types (i.e., creating samples from named distributions) and methods for verifying consistency between different prediction elements.

\subsubsection*{Named prediction elements}
Named prediction elements encode a theoretical probability distribution as a representation of a prediction. 
Only certain distributions are valid for a particular prediction, depending on the target type. E.g., for a discrete, integer-valued target like the number of cases, only distributions for count data (such as Poisson and Negative Binomial) are valid.
A single Named prediction element only takes up one database row.
In practice, a Named prediction can be stochastically converted to a Sample prediction element (using a generator of random deviates from a distribution) and deterministically converted to a Bin prediction element.

\subsubsection*{Sample  prediction elements}

Sample prediction elements are represented by draws from the predictive distribution of interest. 
All samples must be valid data types for the target of interest. E.g., if the target is nominal, with possible outcomes of ``high'', ``moderate'', and ``low'', then every sample must be one of those categories. Or, if the target is continuous with a range of [0,100], then every sample must lie within that range.
Sample forecasts often will occupy the most storage space of any forecast representation (one database row is required for each individual sample). 
However, in the current data model, they can also provide the most information.
Zoltar preserves the order or index of all samples. This means that sample prediction elements across multiple units and targets can be linked by their index, providing an informative joint distribution within a given forecast.

\subsubsection*{Bin  prediction elements}
Bin prediction elements encode an empirical representation of a predictive distribution. They represent all possible realizations or outcomes of target in discrete categories, and each category has an associated probability. 
The probabilities must sum to 1 (+/- a small numerical tolerance).

For nominal and date targets, the valid categories or ``bins'' are required to specified by the user.
For discrete and continuous targets, Bin prediction elements are only accepted if a set of valid categories was specified in the target definition. 
For binary targets, Bin prediction elements may be submitted with a single category of "true".

\subsubsection*{Quantile  prediction elements}
Quantile prediction elements encode a predictive distribution using a set of quantiles of the distribution. Each quantile is represented by a single row of data corresponding to a quantile value in the interval [0,1]. Quantile elements are only valid for continuous, discrete, and date targets.

\subsubsection*{Point  prediction elements}
Point prediction elements consist of a single value, or row, in a database table. 
This value corresponds to the single predicted value from a forecast for a given target-unit-timezero combination.
Point forecasts are deterministic in nature, with no probabilistic structure or uncertainty encoded in them.

\subsubsection*{Data Validation}

Data are validated before being written to the Zoltar database. Validations for individual prediction elements include common-sense checks to ensure that incoming data are of the correct data type and are within the specified range of the target. A complete, updated list of validations is available in the online documentation of the Zoltar system.\cite{zoltarweb}


\subsection*{Forecast scoring}

Many scoring rules for forecast evaluation exist, however the choice of an appropriate metric may depend on the application setting. 
In general, it is desirable for a scoring rule to be `proper'.
Propriety of a scoring rule has a formal mathematical definition, and heuristically means that forecasters are never incentivized to modify their forecast to improve their score.\cite{Gneiting2007} 

As stated in above sections, when a Zoltar project has both forecasts and ground truth data, it automatically computes a set of proper scores for each forecast. 
Scores are updated upon upload of either a new ground truth file or a new forecast. 
We have implemented an asynchronous job queueing system to facilitate the scaling up of computation of many jobs simultaneously.
The speed of scoring depends on the number of jobs in the queue at a given time and the number of available workers to process jobs in the queue.
Scores are computed for unique score/model combinations that are parceled out as individual jobs when new forecasts or truth data are added to a project.
Scores are available for download via the Zoltar API or through the website user interface.\cite{zoltarweb}

\subsection*{Tools to facilitate data access}

To facilitate interaction with data within Zoltar, we have developed two software libraries: zoltr for R and zoltpy for Python.\cite{cornell2020zoltr,cornell2020zoltpy}
These libraries can be used to interact directly with a Zoltar server. 
Using either package, a user can authenticate to the server, query and download existing data about projects, models and forecasts, and upload new forecasts.



\section*{Code availability}

All code for the Zoltar system is available under a GPL-3.0 license from \url{https://github.com/reichlab/forecast-repository/}. Zoltr and zoltpy packages are available under GPL-3.0 licenses from \url{http://reichlab.io/zoltr/} and \url{https://github.com/reichlab/zoltpy}, respectively


\section*{Acknowledgements}

We acknowledge Dr. Wilbert van Panhuis (University of Pittsburgh) and Dr. Michael Johannson (US CDC) for discussions and support throughout the process of designing Zoltar.
This work has been supported by the National Institutes of General Medical Sciences (R35GM119582, PI: Reich) through the Modeling of Infectious Disease Agents Study (MIDAS). The content is solely the responsibility of the authors and does not necessarily represent the official views of NIGMS, MIDAS, or the National Institutes of Health.

\section*{Author contributions}

NGR obtained funding for the project, led the design of Zoltar, zoltr, and zoltpy, and wrote the manuscript.
MC contributed to the design of Zoltar, wrote the code for the system, built the web-based user interface to Zoltar, developed the zoltr and zoltpy software libraries, and provided editorial feedback on the manuscript.
ELR contributed to the design of Zoltar and provided editorial feedback on the manuscript.
KH designed the Zoltar static website, contributed to the zoltr and zoltpy software libraries, and provided editorial feedback on the manuscript.
KL contributed to the zoltr and zoltpy software libraries, and provided editorial feedback on the manuscript.

\section*{Competing interests}
None declared.

\clearpage

\section*{Figures and figures legends}

\clearpage
\section*{Tables}

\begin{table}[h]
\begin{tabular}{lllccccc}
 & &  & \multicolumn{5}{c}{Prediction Elements} \\
target type & data type & obs type  &  point   &  bin   & sample    &  quantile & named  \\
 \hline
 continuous & float & float &   x   &  x   & x & x & x \\
 discrete & int &  int & x   & x & x & x & x \\
 binary & boolean  & boolean & x   &  x   & x & - & - \\
 date & date  & date & x & x & x & x & - \\
 nominal & text & single text value & x & x & x  &  -  & -
\end{tabular}
\caption{This table has one row for each specified type of target, along with the data type associated with that target (for database storage) and ground truth observations of that target. The target\_type field is specified for each target by the user. This specification then unambiguously determines the data\_type and the valid Prediction Elements. Valid prediction elements are indicated by an x in the table cell and invalid prediction elements are indicated by a -.}
\label{tab:target-types}
\end{table}

\begin{table}[]
\begin{tabular}{ll|cc|ccccc}
 & & \multicolumn{2}{c|}{point scores} & \multicolumn{5}{c}{probabilistic scores} \\
target & elem.  &  error  & abs &  log score   & CRPS    &  Brier & PIT & int. score \\
 \hline
continuous & point & x & x & - & x$^a$ & - & - & - \\
& bin & - & - & x & x & x & x  & -\\
& named & - & - & x & x & - & x  & -\\
& sample & - & - & x$^b$ & x & - & x  & - \\
& quantile & - & - & - & - & - & -  & x \\
 \hline
discrete & point & x & x & - & x$^a$ & - & -  & -\\
& bin & - & - & x & x & x & x  & -\\
& named & - & - & x & x & - & x  & -\\
& sample & - & - & x$^b$ & x & - & x  & -\\
& quantile & - & - & - & - & - & -  & x \\
 \hline
nominal & point & - & - & - & - & - & -  & -\\
& bin & - & - & x & - & x & -  & -\\
& sample & - & - & x$^b$ & - & x & -  & -\\
 \hline
binary & point & x & x & - & x$^a$ & - & -  & -\\
& bin & - & - & x & x & x & -  & -\\
& sample & - & - & x$^b$ & x & x & -  & - \\
 \hline
date & point & x & x & - & x$^a$ & - & -  & - \\
& bin & - & - & x & x & x & x  & -\\
& sample & - & - & x$^b$ & x & - & x  & - \\
& quantile & - & - & - & - & - & -  & x 
\end{tabular}
\caption{This table shows the scores available for each combination of target type and prediction element (where these combinations are represented by rows). An `x' indicates that score in the column is available; a `-' indicates that the given score is not available for that particular combination. $a$ denotes that CRPS is equivalent to abssolute error for point forecasts. $b$ denotes that log score is required to be computed by approximation. CRPS and Brier scores are part of a future planned implementation.}
\label{tab:scores}
\end{table}




\clearpage

\bibliographystyle{naturemag}
\bibliography{zoltar}

\end{document}